\documentclass[11pt,twoside]{article}

\begin{document}
\title{\bf Entanglement detection: Linear entropy versus  Bell-CHSH inequality}
\author{Satyabrata Adhikari \thanks{
satyabrata@bose.res.in} \\
 \textsl{S. N. Bose National centre of Basic Sciences, Salt lake, Kolkata 700098, India}\\
}
\date{}
\maketitle{} PACS numbers: 03.67.-a
\begin{abstract}
The relation between the violation of the Bell-CHSH inequalities
and entanglement properties of quantum states is not clear so one
may consider the mixedness of the system to understand the
entanglement properties better than the Bell-CHSH inequality. In
this respect, we prove that if the mixedness of the state measured
by the linear entropy is less than $\frac{2}{3}$ but strictly
greater than zero then the two qubit states are entangled. But if
the linear entropy is greater or equal to $\frac{2}{3}$ then the
state may or may not be entangled. Further we show that our
entanglement criterion detects larger set of entangled state than
Bell-CHSH inequality and Santos's entropic criterion [Phys. Rev. A
\textbf{69}, 022305 (2004)]. Lastly we illustrate our result by
citing few examples.
\end{abstract}

Quantum entanglement is one of the fascinating feature of quantum
mechanics. There is no classical analog of quantum entanglement
and that makes it more fascinating than anything else in physics.
In the field of quantum information theory entanglement plays a
major role. This is also a very useful resource in the sense that
using entanglement one can do many things in the quantum world
which are usually impossible in ordinary classical world. Some of
these tasks are quantum computing \cite{bennett1}, quantum
teleportation \cite{bennett2}, quantum cryptography
\cite{gisin}.\\
The entanglement detection problem is a very genuine and
challenging task in quantum information theory. Researchers accept
this challenge and proposed many entanglement detection methods by
which we could detect the presence of entanglement in a given
system. The first successful candidate was J. S. Bell \cite{bell}
who proposed a entanglement detection scheme (now known as Bell's
inequality), in 1964, when studying the Einstein-Podolsky-Rosen
(EPR) paradox \cite{einstein}. After that many modifications of
the original Bell inequality were proposed. Among all these
Clauser, Horne, Shimony, Holt (CHSH) inequality \cite{chsh} is the
famous one. In 2002, D. Collins et.al. generalizes the Bell-CHSH
inequality for arbitrary d-dimensional systems. This inequality is
popularly known as the Collins-Gisin-Linden-Massar-Popescu
inequality \cite{collins}.
Bell inequalities for multipartite arbitrary dimensional system is also studied \cite{son}.\\
Local realism implies constraints on the statistics of two or more
physically separated systems. These constraints, called Bell
inequalities, can be violated by the statistical predictions of
quantum mechanics. A typical Bell-inequality for bipartite two
qubits system was derived by Clauser, Horne, Shimony, Holt,
allowing more flexibility in local measurement configuration than
the original Bell inequality. The Bell-CHSH inequality read as
\begin{eqnarray}
I_{CHSH}= \langle A_{1}B_{1}\rangle_{\rho}+ \langle
A_{1}B_{2}\rangle_{\rho}+\langle A_{2}B_{1}\rangle_{\rho}-\langle
A_{2}B_{2}\rangle_{\rho} \leq 2\label{bellchsh}
\end{eqnarray}
where $\langle A_{i}B_{j}\rangle_{\rho}=Tr[\rho(\hat{a}_{i}.~
\vec{\sigma}^{A})(\hat{b}_{j}.~\vec{\sigma}^{B})]$ known as the
so-called correlation functions, $\rho$ is the two-qubit state
shared by A and B, $\vec{\sigma}$ is the Pauli matrix vector,
$\hat{a}_{1}$ and $\hat{a}_{2}$ are the unit vectors for the first
and the second measurements performed to the subsystem A
respectively and so do $\hat{b}_{1}$ and $\hat{b}_{2}$ for the
subsystem B. The Bell-CHSH inequality has many merits
\cite{peres}: (i) It is tight, i.e. it defines one of the facets
of the convex polytope of local-realistic (LR) models (ii) It is
violated by all the pure two qubit entangled states (iii) It is
maximally violated by maximally entangled states. Thus the
detection of entanglement in a pure two-qubit system via Bell-CHSH
inequality is totally solved. On the contrary if the given system
is mixed then Bell-CHSH inequality solves the entanglement problem
partially even in two qubit system. The entanglement problem is
not completely solved via Bell-CHSH inequality in case of mixed
state because of the existence of some entangled states which
satisfies the inequality. In this regard Werner \cite{werner}
proposed a class of mixed spin-$\frac{1}{2}$ state which satisfies
the Bell-CHSH inequality although it is entangled. Thus Bell-CHSH
inequality does not detect all entangled state. But if the
inequality is violated by mixed state then the state is surely
entangled. Though Bell-CHSH inequality fails to detect all two
qubits mixed entangled states, it is considered as the most
efficient one because until 2004 there was no example of a quantum
state not violating the CHSH inequality but violating some other
Bell-Inequalities \cite{collins1}. Later And$\dot{a}$s
\cite{andas} showed that a convex combination of product
spin-$\frac{1}{2}$ state does not violate Bell inequalities for
the generalised Bell type observables. Since we are now discussing
about the violation of the Bell-CHSH inequality by a given mixed
state so in this respect we should mention here that it was hard
to say whether a given state violates the CHSH inequality because
one had to construct a respective Bell operator for it. So to make
this hard task easy, Horodecki family provided an effective
criterion (necessary and sufficient condition) for violating the
Bell type inequalities by
mixed spin-$\frac{1}{2}$ state \cite{horodecki1} . The statement of the criterion is as follows:\\
\textbf{Horodecki criterion}\cite{horodecki1}: The two qubit
density matrix $\rho$ violates CHSH inequality for some Bell
operator of the form
$B_{CHSH}=\hat{a}.\vec{\sigma}\otimes(\hat{b}+\hat{b'}).\vec{\sigma}
+\hat{a'}.\vec{\sigma}\otimes(\hat{b}-\hat{b'}).\vec{\sigma}$ iff
\begin{eqnarray}
M(\rho)=max_{i>j}(\lambda_{i}+\lambda_{j})>1 \label{entcond}
\end{eqnarray}
where $\hat{a},~\hat{a'},~\hat{b},~\hat{b'}$ are unit vectors in
$R^{3}$ and $\lambda_{i}'s$ are the eigenvalues of the symmetric
matrix $C_{\rho}^{T}C_{\rho}$(T stands for transposition).\\
Since the relation between the violation of the Bell-CHSH
inequalities and entanglement properties of quantum states is not
clear so one may consider the mixedness of the system to
understand the entanglement properties better than the Bell-CHSH
inequality. It is a known phenomenon that as the mixedness of the
system increases, the entanglement decreases. Naturally a question
arises: Does there exist any upper bound of the mixedness of the
given system upto which the entanglement is stayed in that system
and beyond that the entanglement is totally lost? E. Santos
\cite{santos} studied this problem to some extent and showed that
if the linear entropy, which measures the mixedness of the system,
is less than $\frac{1}{2}$ then there are states which are
entangled. But the bound for the linear entropy given by Santos is
weak in the sense that it only detect entangled states which are
also detected by Bell-CHSH inequality. Bose and Vedral also
studied this type of problem. They gave a lower bound for the
von-Neumann entropy and linear entropy and then showed that if the
entropy exceeds the given bounds then those states cannot be used as a teleportation channel \cite{bose}.\\
In this letter, we provide an upper bound to the mixedness of the
state measured by linear entropy and show that the two qubit
states whose mixedness less than the given upper bound are
entangled. Our result is interesting in the sense that it detects
a larger set of entangled state than any Bell-CHSH inequality and
Santos's entropic criterion. \\\\
\textbf{Theorem:} The two qubit mixed density matrix $\rho$ is
entangled iff
\begin{eqnarray}
S_{L}(\rho)< \frac{2}{3} \label{bnd}
\end{eqnarray}
where $S_{L}(\rho)$ is the linear entropy of the density matrix $\rho$.\\
\textbf{Proof:} Any arbitrary state on $H= C^{2} \otimes C^{2}$
can be represented in a Hilbert-Schmidt basis as follows
\cite{nielsen}:
\begin{eqnarray}
\rho = \frac{1}{4}(I \otimes I + \sum_{i=1}^{3}r_{i}\sigma_{i}
\otimes I + I \otimes \sum_{i=1}^{3}s_{i}\sigma_{i} +
\sum_{i,j=1}^{3} c_{ij} ~ \sigma_{i}\otimes\sigma_{j})
\label{dens.mat.}
\end{eqnarray}
where $I$ represents the $2\times2$ identity matrix,
$\sigma_{i}~~(i=1,2,3)$ denotes the standard Pauli matrices.\\
The coefficients $r_{i}$ and $s_{i}$ are given by
\begin{eqnarray}
r_{i}= Tr(\rho \sigma_{i} \otimes I),~~~~~s_{i}= Tr[\rho(I \otimes
\sigma_{i} )]~~~~~(i=1,2,3) \label{blockvec}
\end{eqnarray}
The coefficients $c_{ij}$ form a real matrix which we call as
$C_{\rho}$ and the elements of the matrix can be evaluated by the
formula
\begin{eqnarray}
c_{ij}= Tr(\rho \sigma_{i} \otimes \sigma_{j})\label{coeff.}
\end{eqnarray}
The state (\ref{dens.mat.}) is pure or mixed according as
$Tr(\rho^{2})=1$ or $Tr(\rho^{2})<1$, where
\begin{eqnarray}
Tr(\rho^{2})=
\frac{1}{4}(1+\sum_{i=1}^{3}r_{i}^{2}+\sum_{i=1}^{3}s_{i}^{2}+\sum_{i,j=1}^{3}c_{ij}^{2})\label{rhosq.}
\end{eqnarray}
For a quantum state $\rho$ in a $4-$ dimensional Hilbert space $H$
the linear entropy is defined as follows \cite{munro}:
\begin{eqnarray}
S_{L}(\rho)= \frac{4}{3}(1-Tr(\rho^{2}))\label{lin.entropy}
\end{eqnarray}
Using eq.(\ref{rhosq.}) and eq.(\ref{lin.entropy}), we get
\begin{eqnarray}
&&S_{L}(\rho)=
1-\frac{1}{3}(\sum_{i=1}^{3}(r_{i}^{2}+s_{i}^{2})+\sum_{i,j=1}^{3}c_{ij}^{2}){}\nonumber\\&&
\Rightarrow M(\rho) \leq Q(\rho)\label{sl}
\end{eqnarray}
where we introduce the function $Q(\rho)$ as
\begin{eqnarray}
Q(\rho)= 3(1-S_{L}(\rho))-\sum_{i=1}^{3}(r_{i}^{2}+s_{i}^{2})
\end{eqnarray}
In eq.(\ref{sl}), we have used the inequality
\begin{eqnarray}
M(\rho)=max_{i>j}(\lambda_{i}+\lambda_{j})\leq
\sum_{i,j=1}^{3}c_{ij}^{2}
\end{eqnarray}
Now to prove the theorem, we have to consider two cases. In the
first case we considered those states which violates the Bell-CHSH
inequality and in the second case we look for those states which
satisfies the Bell-CHSH inequality but violate generalised
Bell-CHSH inequality.\\\\
\textbf{Case-I:} $M(\rho)>1$.\\
In this case the density matrix $\rho$ violates the Bell-CHSH
inequality and hence entangled. Therefore, using Horodecki
criterion and eq. (\ref{sl}), we can say that the state is
entangled iff
\begin{eqnarray}
&&Q(\rho) > 1 {}\nonumber\\&& \Rightarrow S_{L}(\rho)<
\frac{2}{3}-\frac{1}{3}\sum_{i=1}^{3}(r_{i}^{2}+s_{i}^{2})\leq\frac{2}{3}\label{newentcond}
\end{eqnarray}
\textbf{Case-2:} $M(\rho)\leq 1 < Q(\rho)$.\\
In this case the density matrix $\rho$ satisfies the Bell-CHSH
inequality. Since there exist entangled states which satisfies the
Bell-CHSH inequality so it is sufficient to show that the states,
for which the relation
\begin{eqnarray}
M(\rho)\leq 1 < Q(\rho)\label{relation}
\end{eqnarray}
is satisfied, violates the generalised Bell-CHSH inequalities.\\
The generalised Bell-CHSH inequality is violated iff
\cite{horodecki2}
\begin{eqnarray}
N(\rho)>1
\end{eqnarray}
where the function $N(\rho)$ defined as $N(\rho):= Tr \sqrt{C_{\rho}^{T}C_{\rho}}=\sum_{i=1}^{3}\sqrt{\lambda_{i}}$ ~.\\
Recalling the definition of the linear entropy for the state
$\rho$, we have
\begin{eqnarray}
S_{L}(\rho)=
1-\frac{1}{3}(\sum_{i=1}^{3}(r_{i}^{2}+s_{i}^{2})+\sum_{i,j=1}^{3}c_{ij}^{2})\label{linentrcase2}
\end{eqnarray}
The eq. (\ref{linentrcase2}) can be rewritten as
\begin{eqnarray}
Q(\rho)=\sum_{i,j=1}^{3}c_{ij}^{2}=Tr(C_{\rho}^{T}
C_{\rho})=\sum_{i=1}^{3}\lambda_{i}\label{cond.}
\end{eqnarray}
Since $\lambda_{i}\leq 1$ for $i=1,2,3$ \cite{horodecki3} so
\begin{eqnarray}
Q(\rho)=\sum_{i=1}^{3}\lambda_{i}<
\sum_{i=1}^{3}\sqrt{\lambda_{i}}= N(\rho)\label{cond1.}
\end{eqnarray}
Combining eq. (\ref{relation}) and eq. (\ref{cond1.}), we have
\begin{eqnarray}
M(\rho)\leq 1 < Q(\rho) <N(\rho)\label{telrel}
\end{eqnarray}
Therefore, eq. (\ref{telrel}) tells us that the state $\rho$ which
satisfies the relation (\ref{relation}) violates the generalised
Bell-CHSH inequality and hence entangled.\\
Thus the  two qubit state $\rho$ is entangled iff the linear
entropy of the state satisfies $S_{L}(\rho)<
\frac{2}{3}-\frac{1}{3}\sum_{i=1}^{3}(r_{i}^{2}+s_{i}^{2})\leq\frac{2}{3}$.\\
Hence the theorem is proved.\\\\
\textbf{Corollary-1:} The density matrix $\rho$ is useful for
teleportation iff
\begin{eqnarray}
S_{L}(\rho)< \frac{2}{3} \label{extbnd}
\end{eqnarray}
\textbf{Proof:} From eq.(\ref{telrel}), it is clear that the two inequalities $Q(\rho)>1$
and $N(\rho)>1$ simultaneously hold.\\
(a) $Q(\rho)>1\Leftrightarrow S_{L}(\rho)<\frac{2}{3}$
and \\
(b) $N(\rho)>1\Leftrightarrow \rho$ is useful for teleportation.\\
Combining the two logics (a) and (b), we can say that the states
which have linear entropy less than $\frac{2}{3}$ are useful
for teleportation.\\\\
\textbf{Corollary-2:} 
(a) It may happen that there exist states for
which the following inequality holds:
\begin{eqnarray}
Q(\rho) < 1 < N(\rho) \label{qn}
\end{eqnarray}
If inequality (\ref{qn}) holds then there exist states with linear
entropy exceeds $\frac{2}{3}$ useful for teleportation.\\
(b) It may also happen that there exist states for which the
inequality given below holds:
\begin{eqnarray}
Q(\rho) \leq N(\rho) \leq 1 \label{qn1}
\end{eqnarray}
If inequality (\ref{qn1}) holds then there always exist some
separable states with linear entropy exceeds $\frac{2}{3}$ and
hence useless for teleportation.\\
Corollary-2 tells us that there exist states with linear entropy
greater than $\frac{2}{3}$ which may or may not be useful for teleportation.\\
Let us illustrate our results with examples:\\\\
\textbf{Example-1:} Let us consider a two-qubit maximally
entangled mixed state \cite{ishizaka} expressed in the
computational basis as follows:
\begin{eqnarray}
\rho_{MEMS}= \left(\begin{matrix}{\frac{p}{2} & 0 & 0 &
\frac{-p}{2}\cr 0 & 1-p & 0 &0 \cr 0  & 0 & 0 & 0 \cr \frac{-p}{2}
&0 & 0 & \frac{p}{2} }\end{matrix}\right) \label{rhomems}
\end{eqnarray}
where $0\leq p \leq 1$.\\
The real correlation matrix $C_{\rho_{MEMS}}$ for the state
$\rho_{MEMS}$ is given by
\begin{eqnarray}
C_{\rho_{MEMS}}= \left(\begin{matrix}{-p & 0 & 0 & \cr 0 & p & 0
\cr 0  & 0 & 2p-1 }\end{matrix}\right) \label{cormat}
\end{eqnarray}
The eigenvalues of the symmetric matrix
$(C_{\rho_{MEMS}}^{T}C_{\rho_{MEMS}})$ is given by
\begin{eqnarray}
\lambda_{1}= \lambda_{2}= p^{2}, \lambda_{3}=
(1-2p)^{2}\label{eigenvalues}
\end{eqnarray}
The quantity $M(\rho_{MEMS})$ is given by
\begin{eqnarray}
M(\rho_{MEMS})=\left\{\begin{array}{cccc}
2p^{2} & & &\textrm{ when}~~  \frac{1}{3} < p \leq 1\\
5p^{2}-4p+1 & & & \textrm{when}~~ 0 \leq p < \frac{1}{3}
\end{array}
\right. \label{M(rho)1}
\end{eqnarray}
From eq. (\ref{M(rho)1}), it can be easily found out that
\begin{eqnarray}
M(\rho_{MEMS}) \left\{\begin{array}{cccc}
>1 & & &\textrm{ when}~~  \frac{1}{\sqrt{2}} < p \leq 1\\
\leq 1 & & & \textrm{when}~~ 0 \leq p \leq \frac{1}{\sqrt{2}}
\end{array}
\right. \label{ex1entcond}
\end{eqnarray}
Therefore the state $\rho_{MEMS}$ violates the Bell-CHSH inequality when $\frac{1}{\sqrt{2}} < p \leq 1$
and satisfies the inequality when $0 \leq p \leq \frac{1}{\sqrt{2}}$.\\
Now the linear entropy $S_{L}(\rho_{MEMS})$ is given by
\begin{eqnarray}
S_{L}(\rho_{MEMS})=\frac{8p(1-p)}{3}\label{ex1linent}
\end{eqnarray}
The function $\frac{8p(1-p)}{3}$ is symmetric with respect to the
line $p=\frac{1}{2}$. The function is increasing when $0\leq p
\leq \frac{1}{2}$ and decreasing when $\frac{1}{2}\leq p \leq 1$.
The maximum is attained at $p=\frac{1}{2}$. Therefore, we find
that
\begin{eqnarray}
S_{L}(\rho_{MEMS})< \frac{2}{3},~~~~when~~0< p \leq 1
\label{ex1newentcond}
\end{eqnarray}
Using our theorem we conclude that the state $\rho_{MEMS}$ is
entangled for all values of the parameter $p$ except zero. This
may also be verified by calculating the concurrence
\cite{wootters} for the state $\rho_{MEMS}$ which in this case is
found out to be $p$. Therefore the state is separable only when $p=0$, otherwise entangled.\\
Therefore our linear entropic criterion detect entanglement when
$0< p \leq 1$ while Bell-CHSH inequality detect entanglement only
when $\frac{1}{\sqrt{2}} < p \leq 1$. Our criterion thus
detects larger set of entangled states than Bell-CHSH inequality.\\\\
\textbf{Example-2:} In example-1, we find that the maximum value
of the linear entropy $S_{L}$ is $\frac{2}{3}$ when
$p=\frac{1}{2}$. Now it is very likely to consider the states with
$S_{L}>\frac{2}{3}$ because those states may or may not be
entangled. Thus we consider this example to emphasize on the
region $S_{L}>\frac{2}{3}$ by taking a family of Werner state,
which can be expressed in the form:
\begin{eqnarray}
\rho_{W}= \left(\begin{matrix}{\frac{1+r}{4} & 0 & 0 &
\frac{r}{2}\cr 0 & \frac{1-r}{4} & 0 &0 \cr 0  & 0 & \frac{1-r}{4}
& 0 \cr \frac{r}{2} &0 & 0 & \frac{1+r}{4} }\end{matrix}\right)
\label{wernerstate}
\end{eqnarray}
where $0\leq r \leq 1$.\\
The linear entropy for $\rho_{W}$ is given by
\begin{eqnarray}
S_{L}(\rho_{W})=1-r^{2}\label{ex2linent}
\end{eqnarray}
In case of Werner state, Bell-CHSH inequality detect the entangled
state when $r>\frac{1}{\sqrt{2}}$ but our theorem tells us that
the state is entangled iff $r>\frac{1}{\sqrt{3}}$. Therefore,
there is a region $\frac{1}{\sqrt{3}}<r<\frac{1}{\sqrt{2}}$ in
which our theorem detects the entangled state while Bell-CHSH
inequality does not. Thus the condition in terms of linear entropy
given in the theorem is more stronger than the condition (\ref{entcond}).\\
The function $N(\rho)$ for the Werner state can be evaluated in
terms of the parameter $r$ and is given by
\begin{eqnarray}
N(\rho_{W})=3r \label{ex2Nrhocond}
\end{eqnarray}
Therefore the state $\rho_{W}$ is useful for teleportation iff
\begin{eqnarray}
N(\rho_{W})>1 \Rightarrow \frac{1}{3}<r\leq 1 \label{ex2Nrhotel}
\end{eqnarray}
From eq.(\ref{ex2linent}) and eq.(\ref{ex2Nrhotel}), we find that
the state $\rho_{W}$ is useful for teleportation iff $0\leq S_{L}
< \frac{8}{9}$. Thus there exist certain states in the family of
Werner states which are useful in teleportation although their
linear entropy is greater than $\frac{2}{3}$. This example
verifies the first part of the corollary-2. In the region
$\frac{8}{9}\leq S_{L}$, the state $\rho_{W}$ is useless for
teleportation purpose and hence obeying the result obtained by Bose and Vedral.\\\\
To summarize, we have investigated the relation between the two
properties of the two qubit density operator: mixedness and
entanglement. We found that if the linear entropy, which measures
the mixedness of the state, is lying between zero and
$\frac{2}{3}$ (excluding zero) then the two qubit mixed system is
entangled. On one hand, we showed that there exist two qubit
entangled system whose linear entropy greater than $\frac{2}{3}$
and useful in teleportation but on the other hand there are mixed
system with linear entropy greater than $\frac{2}{3}$ are
separable and hence useless in teleportation. Therefore if the
given mixed system has linear entropy greater than $\frac{2}{3}$
then we cannot definitely say that the given system is entangled
or not. Our criterion not only detects entanglement but also
detect larger set of entangled states than Bell-CHSH inequality.
Hence our criterion also shed some light on the detection of two
qubit entanglement. In future we may opt this criterion to detect
the entanglement in d-dimensional and multipartite systems.

\end{document}